\documentclass[aps,prl,twocolumn,showpacs,superscriptaddress]{revtex4}
\usepackage{amssymb}
\usepackage{amsmath}
\usepackage{epsfig}
\usepackage{graphicx}
\usepackage{ulem}

\begin{document}

\title{Interaction-induced connectivity of disordered two-particle states}

\author{D. O. Krimer}
\affiliation{Institute for Theoretical Physics, Vienna University of
Technology, Wiedner Hauptstrasse 8-10/136, 1040 Vienna, Austria}
\author{S. Flach}
\affiliation{New Zealand Institute for Advanced Study, Massey University, Auckland, New Zealand}

\date{\today}

\begin{abstract}
We study the interaction-induced connectivity in the Fock space of two particles in a disordered one-dimensional potential.
Recent computational studies showed that the largest localization length $\xi_2$ of two interacting particles in a weakly random tight binding chain
is increasing unexpectedly slow relative to the single particle localization length $\xi_1$, questioning previous scaling estimates.  We show this to be a consequence of the approximate restoring of momentum conservation of weakly localized single particle eigenstates, and disorder-induced phase shifts for partially overlapping states.
The leading resonant links appear among states which share the same energy and momentum.  We substantiate our analytical approach by computational studies for up to $\xi_1 = 1000$. A potential nontrivial scaling regime sets in for $ \xi_1 \approx 400$, way beyond all previous numerical attacks.
\end{abstract}



\maketitle

{\it Introduction.---} 
The interplay between Anderson localization \cite{anderson1958} and many body interactions \cite{giamarchi1988,aleiner2010}
has been for decades in the research focus of condensed matter. Most theoretical results are not rigorous, and rely on physical intuition, independent computational studies, and of course experimental data. The case of few interacting particles seems to be an exception, as computational approaches are expected to easily do the job here.  For two interacting particles (TIP) in a one-dimensional chain with weak diagonal disorder a number of studies
over the past twenty years produced interesting yet contradicting predictions on the 
scaling of the largest two-particle localization length $\xi_2 \sim \xi_1^{\alpha}$ with the single particle localization length $\xi_1$. These range from $\alpha=2$ \cite{shepelyansky1994,imry1995}, $\alpha=1.6$  \cite{krimer2010} to $\alpha=1$ \cite{schreiber1997,roemer1999}, thus from the existence of a second length scale ($\alpha > 1$) to
the nonexistence of such a scale ($\alpha=1$). Recent computational studies of the TIP eigenstates \cite{krimer2011} show that down to the weakest disorder values accessed by numerical diagonalizations \cite{frahm1995,oppen1996,frahm1999,arias1999}, 
the largest TIP localization length is $\xi_2 \leq 2 \xi_1$ \cite{krimer2011}. Therefore the above scaling predictions are not supported by published numerical results. In another recent study, a surprising TIP wavepacket subdiffusion on
length scales $\xi_1 \ll l \ll \xi_2$ has been found for $\xi_1$ as large as $\xi_1 \approx 400$ 
\cite{Ivanchenko2013}, further fueling the request to understand the TIP dynamics at weak disorder.

In this work we address the intrinsic reasons for the listed discrepancies. 
We focus on the single particle eigenstates (SPE) and compute overlap integrals and connectivities
in the Fock space of two particle eigenstates (TPE) at zero interaction.
We show, that contrary to previous
assumptions,  the overlap integrals show a highly inhomogeneous distribution at weak disorder. SPE gradually restore standing wave phase relations that occur in the tight-binding model without disorder, $W=0$ \cite{Altshuler2003} leading to approximate momentum conservation selection rules in the overlap integrals. At the same time strongly connected TPE have to satisfy approximate energy conservation.
Large connectivities set in at previously unexpected low values of disorder - because
of the combined action of momentum restoring and relative spatial shifts of the SPE on the phase relations between interacting TPE.
We arrive at the surprising conclusion that the rigorous
diagonalization of TIP in the regime of strong connectivity is a matter of future computations, as present
CPUs are hardly capable of doing the job.

{\it Model.---}
We consider the Hubbard Hamiltonian with disorder
\begin{eqnarray}
\label{eq_Hamilt}
{\cal \hat H}\equiv{\cal \hat H}_0+{\cal \hat H}_{int}\;,\; \\
{\cal \hat H}_0 = \sum\limits_l\left[\epsilon_l\hat a_{l}^+\hat a_l+ \hat a_{l+1}^+\hat a_l+\hat
a_{l}^+\hat a_{l+1} \right],\,\, \\
{\cal \hat H}_{int} = \sum\limits_l\left[ \frac{U}{2}\hat a_{l}^+\hat a_l^+\hat a_{l}\hat a_l\right],
\end{eqnarray}
and two indistinguishable bosons.
Note that the results are not changing when considering two distinguishable particles, e.g. two fermions with opposite spins.  

\begin{figure*}
\includegraphics[angle=0,width=.68\columnwidth]{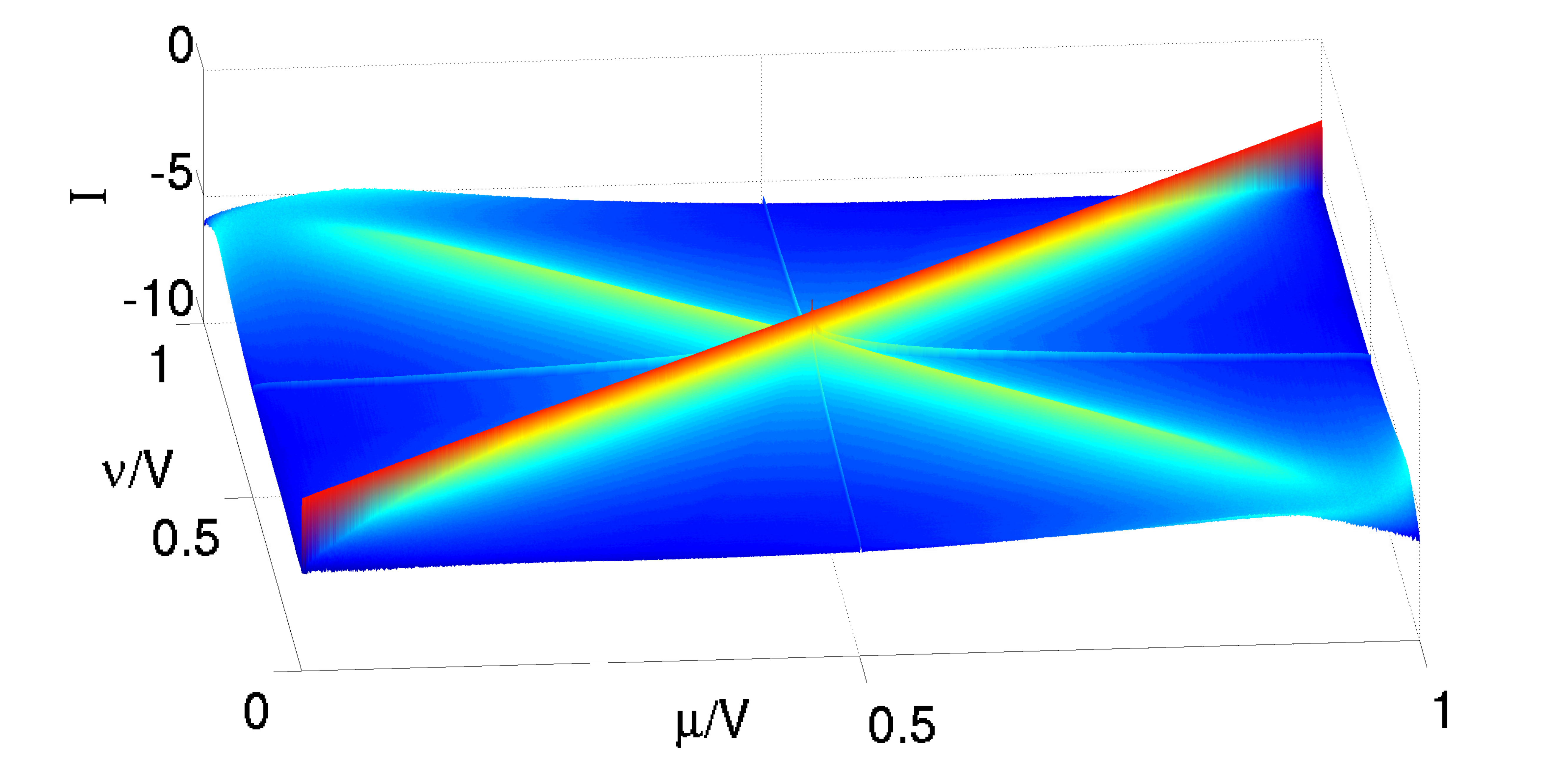}
\includegraphics[angle=0,width=.68\columnwidth]{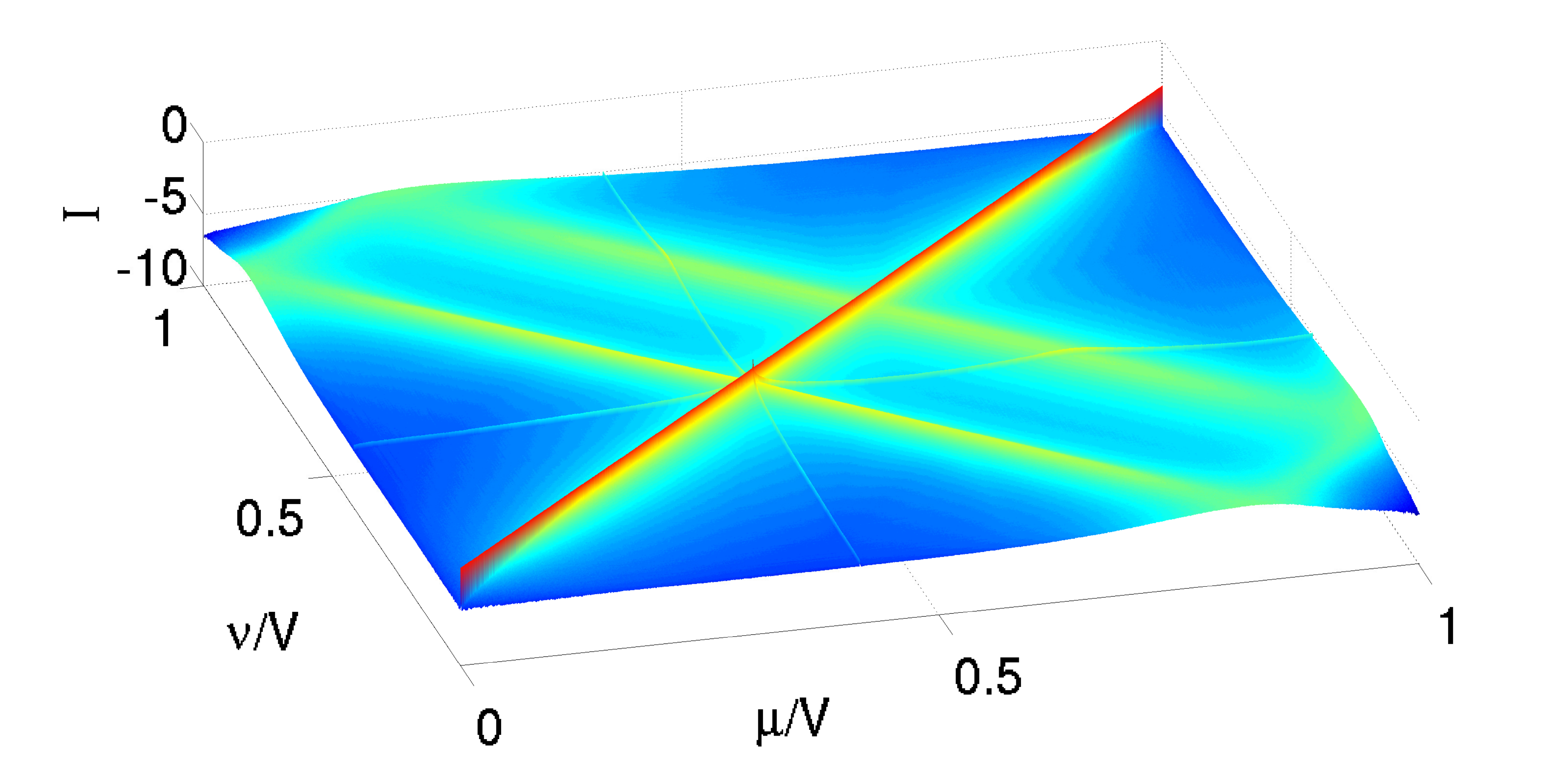}
\includegraphics[angle=0,width=.68\columnwidth]{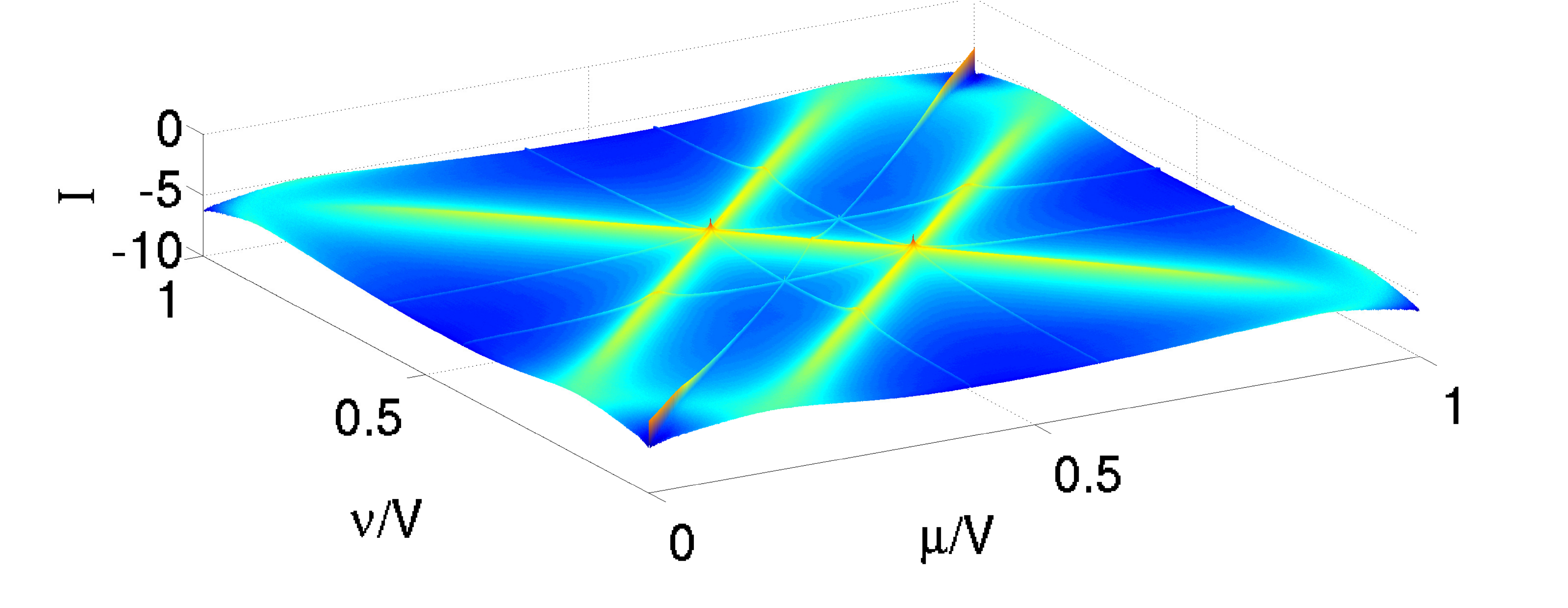}
\caption{Averaged overlap integrals in log scale as a function of the Fock state $|\nu,\mu\rangle$ for the following reference Fock states, from left to right: $\nu_0=\mu_0=V /2$ (center), $\nu_0=\mu_0=V/4 $ (diagonal), $\nu_0/3=\mu_0=V /4$ (antidiagonal). $W=0.5$. }
\label{fig1}
\end{figure*}

The Hamiltonian (\ref{eq_Hamilt}) consists of non-interacting and interacting
parts, ${\cal \hat H}_0$ and ${\cal \hat H}_{int}$, where $\hat a_{l}^+$ and
$\hat a_{l}$ are standard boson creation and annihilation operators on a lattice
site $l$ and $U$ measures the interaction strength. The random uncorrelated
on-site energies $\epsilon_{l}$ are chosen uniformly from the interval
$[-W/2,W/2]$, with $W$ denoting the disorder strength.

{\it One particle.---}
In this case the interaction term does not contribute. Using the basis $|l\rangle\equiv a_{l}^+|0\rangle$ with
$l=1,\ldots ,N$ ($N$ is the number of lattice sites), the SPE  $|\nu\rangle=\sum_{l}^NA_{l}^{(\nu)}| l \rangle$ are defined through the localized eigenvectors $A_l^{(\nu)}\sim e^{-|l|/\xi_1^\nu}$ \cite{anderson1958} of the eigenvalue problem
\begin{equation} \lambda_{\nu} A_l^{(\nu)} = \epsilon_l
A_l^{(\nu)} +(A_{l+1}^{(\nu)} + A_{l-1}^{(\nu)}). \label{AA}
\end{equation}
The eigenvalues $-2-W/2 \leq \lambda \leq 2+W/2$ fill a band with a width $\Delta_1 = 4+W$. The most extended SPE correspond to the band center $\lambda=0$ with localization length
\begin{equation}\label{00}
\xi_1(\lambda=0,W) \approx 100/W^2,
\end{equation}
in the limit of weak disorder $W\leq4$ \cite{kramer1993}. The average volume $V$ which an SPE occupies is estimated to be about $V \approx 3\xi_1$ for weak disorder 
\cite{krimer2010}.

{\it Two particles.---}
For $U=0$ we construct a complete basis of orthonormalized two-particle eigenstates which span a Fock space as product states of SPE 
\begin{equation}
\label{eigen_H0} | \mu,\nu\geq\mu \rangle=\frac{|\mu\rangle\otimes |\nu
\rangle}{\sqrt{1+\delta_{\mu,\nu}}}\;,\; {\cal \hat H}_0 |
\mu,\nu\rangle=(\lambda_\mu+\lambda_\nu)  | \mu,\nu \rangle.
\end{equation}
TIP eigenstates $| q\rangle$ of the interacting particle problem ${\cal \hat H} | q\rangle=\lambda_q |q\rangle$ can be represented in Fock space as $|q\rangle=\sum_{\nu, \mu\le \nu}^N \phi_{\mu\nu}^{(q)} |\mu,\nu \rangle$.  The coefficients $\phi_{\mu\nu}^{(q)}$ satisfy the eigenvalue problem
%
\begin{equation}
\lambda_q\phi_{\mu_0\nu_0}^{(q)}=\lambda_{\mu_0\nu_0}\phi_{\mu_0\nu_0}^{(q)}
+\sum\limits_{\mu,\nu} \frac{2U I^{\mu_0\nu_0}_{\mu\nu}\phi_{\mu\nu}^{(q)}}{\sqrt{1+\delta_{\mu_0\nu_0}}\sqrt{1+\delta_{\mu\nu}}},\label{eq1511}
\end{equation}
%
where
\begin{equation}
I^{\mu_0\nu_0}_{\mu\nu} = \sum_{l}A_l^{(\mu_0)} A_l^{(\nu_0)} A_l^{(\mu)}A_l^{(\nu)}
\label{overlapintegral}
\end{equation}
are the overlap integrals. $\lambda_{\mu_0\nu_0}\equiv \lambda_{\mu_0}+\lambda_{\nu_0}$ and therefore the noninteracting case $U=0$ yields an eigenenergy band with width $\Delta_2=2\Delta_1$ \cite{comment2}.

It follows straight from Eq.(\ref{eq1511}) that two Fock states are strongly (nonperturbatively) coupled
if 
\begin{equation}
R^{\mu_0\nu_0}_{\mu\nu} \equiv \left| \dfrac{\Delta \lambda^{\mu_0\nu_0}_{\mu\nu}}
{U I^{\mu_0\nu_0}_{\mu\nu}}\right| < 1 
\label{R_def}
\end{equation}
where the energy mismatch is given by
\begin{equation}
\Delta \lambda^{\mu_0\nu_0}_{\mu\nu}= \left| \lambda_{\mu_0}+\lambda_{\nu_0} - \lambda_{\mu}-\lambda_{\nu}\right|
\label{DL_def}
\end{equation}
For $U \gg 1$ the interaction separates two particle bound states with double occupancy per site off a continuum of states with one particle per site \cite{krimer2011,winkler2006,scott1994}. In that case, the bound states localize in space even stronger than the single particle states due to the energy separation cost to move one particle.
The remaining states form a Hilbert space of two noninteracting spinless fermions and yield no
increase in the localization length as well (as compared to the single particle case). Therefore,
the strongest effect the interaction can have on increasing the localization length is for $U\approx 1$
which we will assume from here on.

It follows from (\ref{R_def}) that a strong link is realized when the energy mismatch
$\Delta \lambda$ is small (ideally zero) and the overlap integral $I$ is sufficiently large.
The amount of possible strong (resonant) links from a given reference Fock state $|\mu_0,\nu_0\rangle$ is finite. Overlap integrals are exponentially small unless all four single particle states which define
one integral $I$ are sufficiently close to each other in real space. Thus a given reference Fock state has at most approximately 
$V^2$ other basis states which form an interaction network,
from which a resonant subset can be chosen.

{\it Overlap integrals and energy mismatch.---}
We first numerically diagonalize the single particle problem (\ref{AA}). We choose a single particle reference state with energy close to zero, and determine the
subset of all neigbouring SPE in the same localization volume $V$.
We order them with increasing energy corresponding to increasing indices $\nu,\mu$.
The corresponding momentum can be well approximated as $p_{\nu} = \pi \nu / V$.
The obtained two-dimensional momentum space is used to construct interacting Fock states.
Next we choose a reference Fock state with $\nu_0,\mu_0$ for $W=0.5$ and perform a disorder averaging of the overlap integrals $I$
with Fock states with some given $\nu,\mu$.
The result is shown in Fig.\ref{fig1} for the reference state being at the center, the diagonal,
and the antidiagonal of the two-dimensional momentum space.

We find that the overlap integrals are predominantly nonzero along certain straight lines. These lines 
follow
simple momentum conservation rules for two interacting particles in a box of size $V$ \cite{pinto2009}, underpinning therefore that $\nu$ is a momentum index for weak disorder.
This is one of the reasons why previous attempts to estimate averages of overlap integrals over the whole momentum space were not useful \cite{shepelyansky1994,imry1995}.

Let us minimize the energy mismatch (\ref{DL_def}). The single particle
energy can be estimated as $-2\cos (p) = -2\cos \pi \frac{\nu}{V}$. 
Therefore the energy mismatch is exactly zeroed if the condition
\begin{equation}
\cos \pi \frac{\nu_0}{V} + \cos \pi \frac{\mu_0}{V} = \cos \pi \frac{\nu}{V} + \cos \pi \frac{\mu}{V}
\label{mismatch}
\end{equation}
is satisfied. It defines some curved line in $\{\nu,\mu\}$ space.
The notable exception is the antidiagonal
straight
line in Fig.\ref{fig1} (left and right plots) which does conserve both the energy and the momentum.
 Note that this coincidence of momentum and energy conservation for
pairs of two particle Fock states along the antidiagonal is the result of the restoring of a particle hole symmetry of the considered model in the limit of vanishing disorder. 
The tight binding model is a member of a family of models with hopping over
odd distances in real space only, which allows the introduction of AB sublattices, and
results (for $W=0$) in the eigenvector property $A_l(\lambda) = (-1)^l A_l(-\lambda)$ \cite{Altshuler2003}.

\begin{figure}
\includegraphics[angle=0,width=.6\columnwidth]{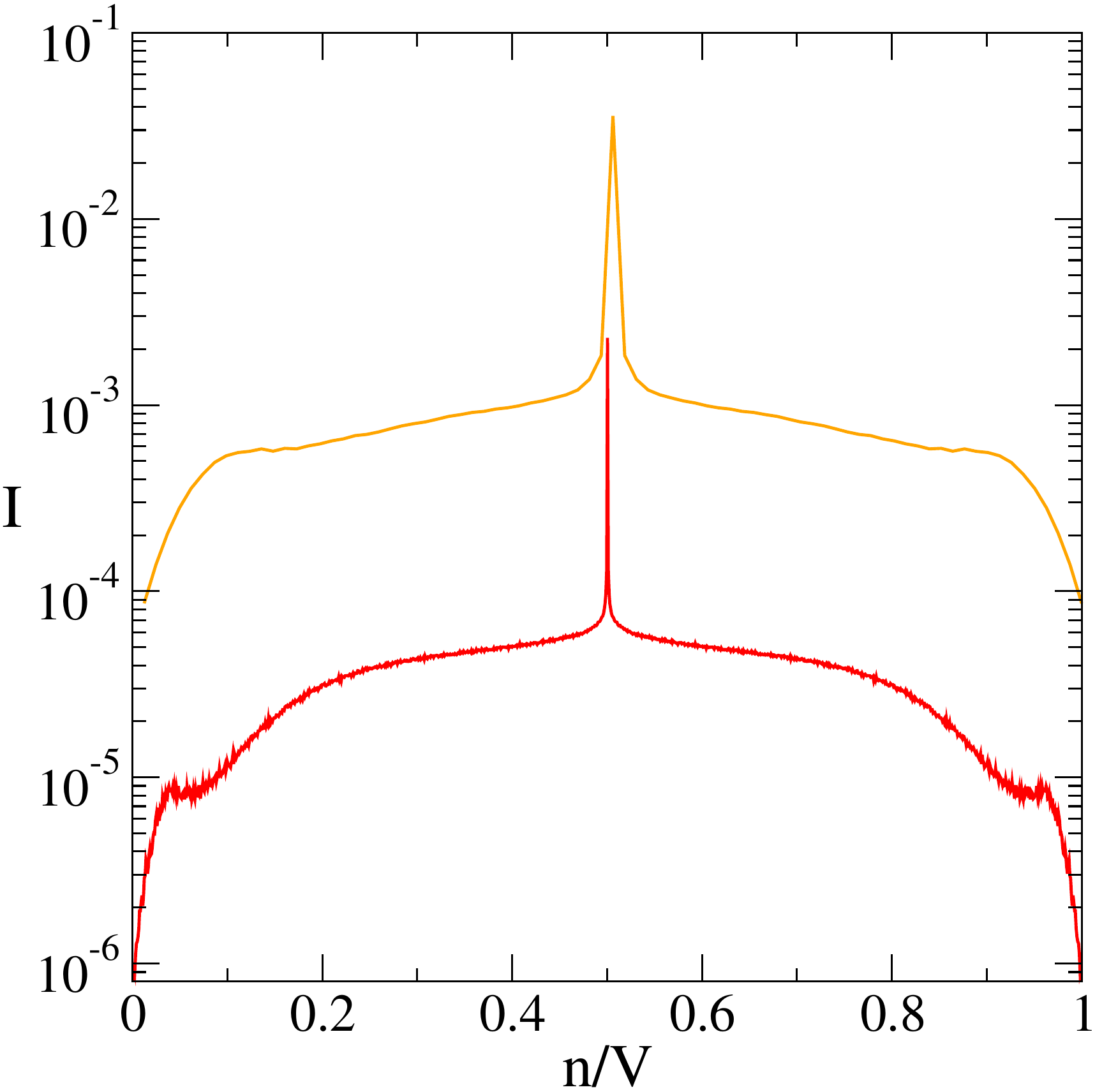}
\caption{Averaged overlap integrals along the antidiagonal for $W=2$ (upper curve) and $W=0.5$ (lower curve). The reference Fock state is at the center $\nu_0=\mu_0=V/2$.}
\label{fig2}
\end{figure}

We focus on the subset of Fock states
along the antidiagonal only with
$\nu_0=\mu_0=V/2$
(center) and $\nu + \mu = \nu_0 + \mu_0$.
It is this tiny subset which is capable to set up the strongest
resonant network and substantially delocalize two interacting particles, as compared to one.
We plot the variation of the overlap integrals along the antidiagonal for $W=2$ and
$W=0.5$ in Fig.\ref{fig2}.
We observe a peak at the center which corresponds to $\langle I_{\nu_0\nu_0}^{\nu_0\nu_0}\rangle $. Its value can be estimated using normalization properties of SPE
as $\langle I_{\nu_0\nu_0}^{\nu_0\nu_0}\rangle \approx 1/V $. For $W=2$ this yelds 0.013 and for $W=0.5$ respectively 0.0008, which are reasonably
close to the numerical data 0.03 and 0.002. In particular their ratio is 16 from the estimate and 15 from numerics, showing that we correctly determine the scaling.
Off the peak we find a plateau at significantly reduced values $\sim 10^{-3}$ ($W=2$) and 
$\sim 10^{-5}$ ($W=0.5$). This reduction is due to relative shifts of SPE in real space. For weak disorder, assume
that each SPE is given by $A^{(\nu)}_{l}=\frac{1}{\sqrt{V}} {\rm e}^{2\pi i (l-l_{\nu})\nu/V}$ for
$l_{\nu} \leq l \leq  l_{\nu}+V$, where $l_{\nu}$ encodes the spatial position of the SPE. The average
of the overlap integral along the antidiagonal  is equivalent to averaging
$I^{V/2,V/2}_{\nu,V-\nu}$ over $\nu$ and over all possible values of $l_{\nu},l_{\mu}$.  This yields
$\langle I_{ad} \rangle = \frac{3}{4V^2}$. The numerical prefactor $3/4$ originates from the relative
shift of flat and constant distributions along the lattice. The scaling $1/V^2$ however is due to the phase mismatch of SPE shifted
relatively to each other \cite{comment3}. This scaling is much weaker than the $V^{-3/2}$ law predicted in
\cite{shepelyansky1994,imry1995} because the standing wave phase correlations were neglected.
The distributions of the energy mismatch $\Delta \lambda$ along the antidiagonal follow approximately 
a normal distribution with the characteristic energy scale $\Delta_1$ \cite{flach2014}, due to the central limit theorem already at work. A quick estimate of the probability of resonance (\ref{R_def}) yields a number independent of
$V$. Therefore fluctuations of the overlap integral values, and their correlations to the energy mismatch 
might be of decisive importance.

\begin{figure}
\includegraphics[angle=0,width=\columnwidth]{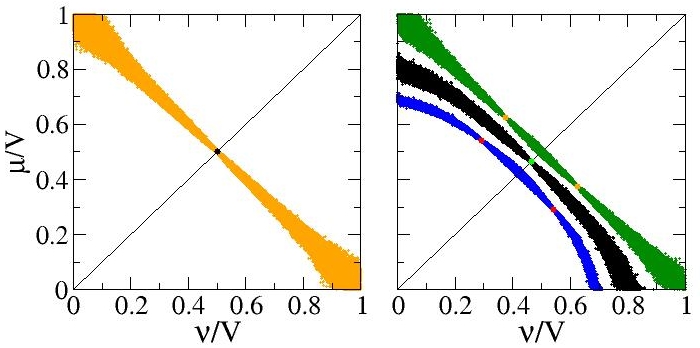}
\caption{Points represent those pairs of $\mu$, $\nu$ which satisfy the resonance condition $R^{\mu_0\nu_0}_{\mu\nu} \le 1$ with the reference modes $\mu_0$, $\nu_0$. Calculations are performed for $W=0.5$ and different pairs of $\mu_0$, $\nu_0$ (shown by filled circles) located at the centre (left panel) and at antidiagonal, diagonal and arbitrary location of $\mu_0$, $\nu_0$ (right panel).}
\label{fig3}
\end{figure}

Let us turn to numerical data.
In Fig.\ref{fig3} we show the observed locations of all resonant partner Fock states ($R < 1$) for different reference Fock states. 
For a reference state
$\nu_0=\mu_0=V/2$ we nicely observe the grouping of all network partners along the antidiagonal (Fig.\ref{fig3} left plot). For another reference state $\nu_0=0.4 V\;,\; \mu_0 = 0.6 V$ on the antidiagional the network partners
still belong to the antidiagonal neighborhood, simply their number decreases (Fig.\ref{fig3} right plot). 
For reference states off the antidiagonal (Fig.\ref{fig3} right plot) the network partners are located close to curved lines
which are a manifestation of the single particle dispersion (energy conservation, see Eq. (\ref{mismatch})), with even smaller partner numbers. Therefore we confirm that resonances are defined by momentum and energy conservation.

\begin{figure}
\hspace*{-1.5cm}
\includegraphics[angle=-0,width=1.\columnwidth]{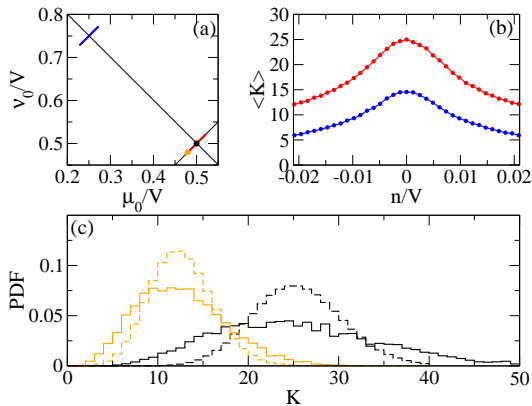}
\caption{Statistical properties of the coordination number $K$. (a) The colored line segments perpendicular to the
antidiagonal represent  the pairs of $\mu_0$, $\nu_0$ taken in the calculations. (b) average
coordination number $\langle K \rangle$ (solid lines with symbols) for the reference modes from the line
segments in (a) (the same coloring is kept). Here $n/V$ measures the deviation along the line segments from
the antidiagonal.
(c) PDFs ${\cal W}(K)$ for two pairs of $\mu_0$ and $\nu_0$. Dashed lines: corresponding Binomial distributions with the same average
coordination number $\langle K\rangle$. The strength of disorder $W=0.35$.}
\label{fig4}
\end{figure}

{\it Connectivity.---}
A central property of any network is the connectivity $K$ - the number of links from a given
reference state to other partners. 
Values of $\langle K \rangle \leq 2$ do not lead to any substantial increase of the localization
length, as there is a high probability to terminate the path after one or two connections.
For a given pair $\mu_0$ and $\nu_0$ the connectivity $K$ to all pairs of modes $\mu$ and $\nu$ residing the same localization volume, is defined as the number of connections for which $R$ fulfills the condition (\ref{R_def}) (if any). We evaluated its average $\langle K \rangle$ for $\nu_0=\mu_0=V/2$:
$\langle K \rangle (W=2)=1.1$, $\langle K \rangle (W=0.75)=5.4$, $\langle K \rangle (W=0.5)=12$,
$\langle K \rangle (W=0.35)=25$. Therefore the potentially interesting regime of sufficiently large
connectivity is accessed only for $W < 1$. All previous diagonalization studies were exploring $1 < W < 4$.
Despite the fact that $V(W=4) \approx 20$ and $V(W=1) \approx 300$, the numerically accessed parameter interval turns out to be irrelevant for the study of a possible dramatic increase in $\xi_2$. 
The reason is the above discussed smallness in the overlap integrals which originates from the 
relative shifts of SPE with standing wave phase relations.
We note that the connectivity increase for $W < 1$ happens also for other choices of $\mu_0,\nu_0$.
For instance for $\nu_0=V/4$, $\mu_0 = 3V/4$ we find  $\langle K \rangle (W=0.5)=7.5$.
Still it is much weaker than the numbers obtained for the antidiagonal, as shown in Fig.\ref{fig4}(a,b).
In Fig.\ref{fig4}(c) we show the distribution of $K$ for $W=0.35$ and two reference Fock states - for the center and the diagonal in Fig.\ref{fig1}.
%

Larger values of $\langle K \rangle$ do not necessarily lead to an enhancement 
of the localization length, since there can be closed loops in the resonant network in Fock space,
whose length simply increases. The existence of loops is enforcing a certain degree of
correlations between the resonant links. Assume the opposite - i.e. that the actual values of 
$R < 1$ which define a set of links from a given reference state to other Fock states, are 
not correlated. 
Then, the connectivity $K$ must be binomially distributed because $K$ would be nothing more than a number of successful events $R<1$ in a sequence of $L$  independent yes/no experiments, each of which yields success with probability $p$. Thus one expects
\begin{equation}
{\cal W}(K)=\dfrac{L!}{k!(L-k)!}p^K(1-p)^{L-K}.
\end{equation}
Note, that the average coordination number $\langle K \rangle$ for the binomial distribution is
related to $p$ as $\langle K \rangle=L\cdot p$. On the other hand, we know $\langle K \rangle$ from
the numerical simulations and therefore can easily calculate the success probability $p$. We test
this hypothesis. Results of the comparison of the numerically obtained  ${\cal W}(K)$ and corresponding binomial
distributions are shown in Fig.~\ref{fig4}(c). We observe a strong deviation of numerical PDFs from the binomial distributions, concluding that resonances are not completely independent events. This
might be a hint 
that resonant loops in Fock space are formed, which could act against delocalization.



\section{Conclusions}

We have shown, that contrary to previous assumptions,  a possible substantial increase in the localization length of two interacting particles in a random potential sets in at unexpectedly weak disorder values. This is due to a gradual restoring of momentum conservation in single particle eigenstates in the limit of vanishing disorder.
That in turn enforces a highly inhomogeneous resonance network of matrix elements. The scaling of the overlap integrals along the resonant network is much weaker than predicted in previous papers because phase correlations and relative position shifts of eigenstates
have to be taken into account. Resonant links between Fock states follow the resonance network.
The connectivity in Fock space  grows substantially with weak disorder, indicating the possibility 
of the emergence of a new localization ength scale for two interacting particles. Because this potential
regime is setting in at anomalously weak disorder strength, previous numerical scaling tests are
not conclusive (too strong disorder). But even more, with current computers and exact diagonalization
methods it is highly nontrivial to enter the desired potential scaling regime which starts at $W=0.5$ and should
extend at least down to $W=0.05$ to estimate exponents. Therefore we are in need of new computational methods.
We also conjecture that a breaking of particle-hole symmetry by adding next-to-nearest neighbour hoppings will lead to a further supression of the delocalization trend by interactions.

\section{Acknowledgements}
We thank Boris Altshuler and Igor Aleiner for many useful and fruitful discussions.


\begin{thebibliography}{99}
\bibitem{comment2} 
For numerical purposes we expand the two particle eigenstates $|q\rangle$ in the local basis $|q\rangle=\sum_{m, l\le m}^N{\cal L}_{l,m}^{(q)}| l,m \rangle, \quad  |l,m \rangle\equiv a_{l}^+a_{m}^+|0\rangle/(\sqrt{1+\delta_{lm}})$, where ${\cal L}_{l,m}^{(q)}=\langle l,m | q\rangle$ are normalized eigenvectors and $|0\rangle $ is the vacuum state.
\bibitem{comment3}
A similar calculation for SPE which are not shifted relative to each other will give $\langle I_{ad} \rangle 
= 1/V$.
\end{thebibliography}
\end{document}